\documentclass[12pt]{article}
\usepackage{amssymb,amsmath,epsfig}
\begin{document} \parindent=0pt
\parskip=6pt \rm

\begin{center}

{\bf \Large On the phase diagrams of the ferromagnetic
superconductors UGe$_2$ and ZrZn$_2$}

\vspace{0.3cm}

{\bf Michael G. Cottam$^1$, Diana V. Shopova$^2$, and Dimo I.
Uzunov$^{1,2\ast}$}

$^1$ Department of Physics and Astronomy, University of Western
Ontario, London, Ontario N6A 3K7, Canada.\\
$^2$ CP Laboratory, Institute of Solid State Physics, Bulgarian
Academy of Sciences, BG-1784 Sofia, Bulgaria.\\
$^{\ast}$ Corresponding author. Email: d.i.uzunov@gmail.com.
\end{center}

{\bf Pacs}: 74.20.De, 74.25.Dw, 64.70.Tg\\



\begin{abstract}

A general phenomenological theory is presented for the phase
behavior of ferromagnetic superconductors with spin-triplet
electron Cooper pairing. The theory accounts in detail for the
temperature-pressure phase diagram of ZrZn$_2$, while the main
features of the diagram for UGe$_2$ are also described.
Quantitative criteria are deduced for the U-type (type I) and
Zr-type (type II) unconventional ferromagnetic superconductors
with spin-triplet Cooper electron pairing. Some basic properties
of quantum phase transitions are also elucidated.

\end{abstract}

\vspace{0.5cm}

The remarkable coexistence of itinerant ferromagnetism and
unconventional (spin-triplet) superconductivity at low
temperatures ($T < 1$ K) was discovered experimentally in the
intermetallic compounds UGe$_2$ \cite{Saxena:2000, Tateiwa:2001,
Harada:2007}, ZrZn$_2$ \cite{Pfleiderer:2001}, and URhGe
\cite{Aoki:2001}. Other metallic compounds, such as UCoGe
\cite{Huy:2007, Huy:2008} and UIr \cite{Akazawa:2005,
Kobayashi:2006}, were also found to be spin-triplet ferromagnetic
superconductors. In ZrZn$_2$, URhGe, and UCoGe, the mixed phase of
coexistence of ferromagnetism and unconventional superconductivity
(labeled the FS phase) occurs over a wide range of pressure (i.e.,
from ambient pressure $P \sim 1$ bar up to $10$ kbar). By
contrast, in other compounds (e.g., UGe$_2$ and UIr) this FS phase
is found only in the high-pressure part ($P\sim 10$ kbar) of the
$T-P$ phase diagram.

Another feature of the above compounds is that the FS phase occurs
only in the ferromagnetic phase domain of the $T-P$ diagram.
Specifically, at equilibrium and a given $P$, the temperature
$T_{F}(P)$ of the normal-to-ferromagnetic phase (or N-FM)
transition is never lower than the temperature $T_{FS}(P)$ of the
ferromagnetic-to-FS phase (or FM-FS) transition. This is
consistent with the point of view that the superconductivity in
these compounds is triggered by the spontaneous magnetization
$\mbox{\boldmath$M$}$, by analogy with the well-known triggering
of the superfluid phase A$_1$ in $^3$He at mK temperatures by the
external magnetic field $\mbox{\boldmath$H$}$. This helium-analogy
has been used in some theoretical studies (see, e.g.,
\cite{Machida:2001, Walker:2002, Shopova:2003, Shopova:2005}),
where Ginzburg-Landau (GL) free energy terms to describe the FS
phase were derived by symmetry arguments.

For the spin-triplet ferromagnetic superconductors the trigger
mechanism was recently examined in detail \cite{Shopova:2005}. The
main system properties are affected by a term in the GL expansion
of the form $\mbox{\boldmath$M$} |\mbox{\boldmath$\psi$}|^2$,
which represents the interaction of $\mbox{\boldmath$M$} = \{M_j;
j=1,2,3\}$ with the complex superconducting vector field
$\mbox{\boldmath$\psi$} =\{\psi_j\}$. Specifically, this term
triggers $\mbox{\boldmath$\psi$} \neq 0$ for certain $T$ and $P$
values. An analogous trigger mechanism is familiar in the context
of improper ferroelectrics \cite{Cowley:1980}.

A crucial consideration in this work is the nonzero magnetic
moment of the spin-triplet Cooper pairs of the electrons. While
the spin-singlet Cooper pairs have net spin zero and are quite
sensitive to the magnitude of the magnetic induction
$\mbox{\boldmath$B$}$, the spin-triplet pairs are known to be
robust with respect to relatively large $\mbox{\boldmath$B$}$. The
phenomena of spin-triplet superconductivity and itinerant
ferromagnetism are both due to the same electron bands of the
compounds: the $f$-band electrons in uranium-based compounds and
the $d$-band electrons in ZrZn$_2$. However, the microscopic band
theory of magnetism and superconductivity in non-Fermi liquids of
strongly interacting heavy electrons is either too complex or
insufficiently developed to describe the complicated behavior in
itinerant ferromagnetic compounds. Consequently, several authors
(see \cite{Machida:2001, Walker:2002, Shopova:2003, Shopova:2005,
Linder:2008}) have explored a phenomenological description within
self-consistent mean field theory, and we build on a similar
approach here.

In this Letter, by focusing on ZrZn$_2$ and UGe$_2$ with their
contrasting types of behavior, we show that the $T-P$ phase
diagrams of spin-triplet ferromagnetic superconductors can be
successfully described starting from the general GL free energy
$F(\mbox{\boldmath$\psi$},\mbox{\boldmath$M$})$ established in
Refs. \cite{Machida:2001, Walker:2002, Shopova:2003,
Shopova:2005}. The present phenomenological approach includes both
mean-field and spin-fluctuation theory (SFT), as in
\cite{Yamada:1993}, considerations. We propose a simple, yet
comprehensive, modeling of the $P$ dependence of the free energy
parameters, from which it is shown that the phase diagram of
ZrZn$_2$ is obtained in good quantitative agreement with the
experimental data~\cite{Pfleiderer:2001}. Further, the main
features~\cite{Saxena:2000} of the $T-P$ diagram of UGe$_2$ are
also well-described within our approach. The theory is capable of
outlining several different possible topologies for the $T-P$
phase diagram, depending on the GL parameters of the material
which can be chosen in accordance with experiment. Quantitative
criteria emerge for two distinct types of behavior for
unconventional ferromagnetic superconductors, which we label
Zr-type and U-type. Further possible applications are to URhGe,
UCoGe, and UIr. Our results address questions regarding the order
of the quantum phase transitions at ultra-low and zero
temperatures. They also pose intriguing questions pointing to
further experimental investigations of (e.g.) the detailed
structure of the phase diagrams in the high-$P$/low-$T$ region.

Following Ref. \cite{Shopova:2005} the free energy per unit
volume, $F/V =f(\mbox{\boldmath$\psi$},\mbox{\boldmath$M$})$, can
be written in the form

\begin{equation}
\label{Eq1} f(\mbox{\boldmath$\psi$},\mbox{\boldmath$M$})=
  a_s|\mbox{\boldmath$\psi$}|^2 +\frac{b_s}{2}|\mbox{\boldmath$\psi$}|^4 +
  \frac{u_s}{2}|\mbox{\boldmath$\psi$}^2|^2 +
\frac{v_s}{2}\sum_{j=1}^{3}|\psi_j|^4 +
 a_f\mbox{\boldmath$M$}^2 +
 \frac{b_f}{2}\mbox{\boldmath$M$}^4 + i\gamma_0 \mbox{\boldmath$M$}
 \cdot (\mbox{\boldmath$\psi$}\times \mbox{\boldmath$\psi$}^*) + \delta
\mbox{\boldmath$M$}^2 |\mbox{\boldmath$\psi$}|^2.
\end{equation}

\noindent The material parameters satisfy, as in
\cite{Shopova:2005}, $b_s >0$, $b_f>0$, $a_s = \alpha_s(T-T_{s})$,
and $a_f = \alpha_{f}[T^n-T_{f}^n(P)]$, where $n=1$ gives the
standard form of $a_f$, and $n=2$ applies for
SFT~\cite{Yamada:1993}. The terms proportional to $u_s$ and $v_s$
describe, respectively, the anisotropy of the spin-triplet
electron Cooper pairs and the crystal anisotropy. Next, $\gamma_0
\sim J$ (with $J>0$ the ferromagnetic exchange constant) and
$\delta > 0$ are parameters of the
$\mbox{\boldmath$\psi$}$-$\mbox{\boldmath$M$}$ interaction terms.
Previous mean-field studies have shown that the anisotropies
represented by the $u_s$ and $v_s$ terms in Eq. (\ref{Eq1})
slightly perturb the size and shape of the stability domains of
the phases, while similar effects can be achieved by varying the
$b_s$ factor in the $b_s|\mbox{\boldmath$\psi$}|^4$ term. For
these reasons, in the present analysis we ignore the anisotropy
terms, setting $u_s = v_s = 0$, and consider $b_s\equiv b >0$ as
an effective parameter. Then, without loss of generality, we are
free to choose the magnetization vector to have the form
$\mbox{\boldmath$M$} = (0,0,M)$.

A convenient dimensionless free energy can now be defined by
$\tilde{f} = f/(b_f M_0^4)$, where $M_0 = [\alpha_fT_{f0}^n
/b_f]^{1/2} >0$ is the value of $M$ corresponding to the pure
magnetic subsystem $(\mbox{\boldmath$\psi$} \equiv 0)$ at $T=P=0$
and $T_{f0}=T_f(0)$. On scaling the order parameters as $m =
M/M_0$ and $\mbox{\boldmath$\varphi$} = \mbox{\boldmath$\psi$}
/[(b_f/b)^{1/4}M_0]$ we obtain

\begin{equation}
\label{Eq2} \tilde{f}= r\phi^2 + \frac{\phi^4}{2}+ tm^2
+\frac{m^4}{2} + 2\gamma m\phi_1\phi_2\mbox{sin}\theta +
\gamma_1m^2\phi^2,
\end{equation}

\noindent where $\phi_j =|\varphi_j|$, $\phi =
|\mbox{\boldmath$\varphi$}|$, and $\theta$ is the phase angle
between the complex $\varphi_2$ and $\varphi_1$. The dimensionless
constants are $t = [\tilde{T}^n-\tilde{T}_f^n(P)]$, $r = \kappa
(\tilde{T}-\tilde{T}_s)$ with $\kappa =
\alpha_sb_f^{1/2}/\alpha_fb^{1/2}T_{f0}^{n-1}$, $\gamma =
\gamma_0/ [\alpha_fT_{f0}^nb]^{1/2}$, and $\gamma_1 =
\delta/(bb_f)^{1/2}$. The reduced temperatures are $\tilde{T} =
T/T_{f0}$, $\tilde{T}_f(P) = T_f(P)/T_{f0}$, $\tilde{T}_s(P)=
T_s(P)/T_{f0}$.

The analysis involves making simple assumptions for the $P$
dependence of the $t$, $r$, $\gamma$, and $\gamma_1$ parameters in
Eq. (\ref{Eq2}). Specifically, we assume that only $T_f$ has a
significant $P$ dependence, described by $\tilde{T}_f(P) = (1 -
\tilde{P})^{1/n}$, where $\tilde{P} = P/P_0$ and $P_0$ is a
characteristic pressure deduced later. In ZrZn$_2$ and UGe$_2$ the
$P_0$ values are very close to the critical pressure $P_c$ at
which both the ferromagnetic and superconducting orders vanish, but
in other systems this is not necessarily the case. As we will
discuss, the nonlinearity ($n=2$) of $T_f(P)$ in ZrZn$_2$ and
UGe$_2$ is relevant at relatively high $P$, at which the N-FM
transition temperature $T_F(P)$ may not coincide with $T_f(P)$.

The simplified model in Eq. (\ref{Eq2}) is capable of describing
the main thermodynamic properties of spin-triplet ferromagnetic
superconductors. There are three stable phases: (i) the normal (N)
phase, given by $\phi = m = 0$; (ii) the pure ferromagnetic (FM)
phase, given by $m = (-t)^{1/2} > 0$, $\phi =0$; and (iii) the FS
phase, given by $\phi_1^2= \phi_2^2= ( \gamma m-r-\gamma_1
m^2)/2$, $\phi_3 = 0$, where $\mbox{sin}\theta = -1$ and $m$
satisfies

\begin{equation}
\label{Eq3} (1-\gamma_1^2) m^3 + \frac{3}{2} \gamma \gamma_1 m^2
+\left(t-\frac{\gamma^2}{2}-\gamma_1 r\right) m + \frac{\gamma
r}{2}=0.
\end{equation}

\noindent We note that FS is a two-domain phase as discussed in
Ref. \cite{Shopova:2003, Shopova:2005}. Although Eq. (\ref{Eq3})
is complicated, some analytical results follow, e.g., we find that
the second order phase transition line $\tilde{T}_{FS}(P)$
separating the FM and FS phases is the solution of

\begin{equation}
\label{Eq4} \tilde{T}_{FS}(P) = \tilde{T}_s +
\frac{\gamma_1}{\kappa}t(T_{FS}) +
\frac{\gamma}{\kappa}[-t(T_{FS})]^{1/2}.
\end{equation}

Under certain conditions, the $T_{FS}(P)$ curve has a maximum at
$\tilde{T}_{m} = \tilde{T}_s +(\gamma^2/4\kappa\gamma_1)$ with
pressure $P_m$ found by solving $t(T_m,P_m)=
-(\gamma^2/4\gamma_1^2)$. Examples will be given later, but
generally this curve extends from ambient $P$ up to a tri-critical
point labeled B, with coordinates $(P_B,T_B)$, where the FM-FS
phase transition occurs at a straight line of first order
transition up to a critical-end point C. The lines of all three
phase transitions (N-FM, N-FS, and FM-FS) terminate at C. For $P >
P_C$ the FM-FS phase transition occurs on a rather flat, smooth
line of equilibrium transition of first order up to a second
tricritical point A with $P_A \sim P_0$ and $T_A \sim 0$. Finally,
the third transition line terminating at C describes the second
order phase transition N-FM. The temperatures at the three
multi-critical points correspond to $\tilde{T}_A = \tilde{T}_s$,
$\tilde{T}_B = \tilde{T}_s + {\gamma^2(2
+\gamma_1)}/{4\kappa(1+\gamma_1)^2}$, and $\tilde{T}_C =
\tilde{T}_s + {\gamma^2}/{4\kappa(1+\gamma_1)}$, while the $P$
values can be deduced from the previous equations. These results
are valid whenever $T_f(P) > T_s(P)$, which excludes any pure
superconducting phase ($\mbox{\boldmath$\psi$} \neq 0, m = 0$) in
accord with the available experimental data. Note that, for any
set of material parameters, $T_A < T_C < T_B < T_m$ and
$P_m<P_B<P_C$.


A calculation of the $T-P$ diagram from Eq. (\ref{Eq2}) for any
material requires some knowledge of $P_0$, $T_{f0}$, $T_s$,
$\kappa$ $\gamma$, and $\gamma_1$. The temperature $T_{f0}$ can be
obtained directly from the experimental phase diagrams. The model
pressure $P_0$ is either identical to or very close to the
critical pressure $P_c$ at which the N-FM phase transition line
terminates at $T \sim 0$. The characteristic temperature $T_s$ of
the generic superconducting transition is not available from the
experiments and thus has to be estimated using general consistency
arguments. For $T_f(P)> T_s(P)$ we must have $T_s(P) =0$ at $P
\geq P_c$, where $T_f(P) \leq 0$. For $0 \leq P\leq P_0$, $T_s <
T_C$ and therefore for cases where  $T_C$ is too small to be
observed experimentally, $T_s$ can be ignored. For systems where
$T_C$ is measurable this argument does not hold. This is likely to
happen for $T_s > 0$ (for $T_s < 0$, $T_C$ is very small).
However, in such cases, pure superconducting phase should be
observable. To date there are no experimental results reported for
such a feature in ZrZn$_2$ or UGe$_2$, and thus we can put $T_s =
0$. We remark that negative values of $T_s$ are possible, and they
describe a phase diagram topology in which the FM-FS transition
line terminates at $T=0$ for $P< P_c$. This might be of relevance
for other compounds, e.g., URhGe.

Typically, additional features of the experimental phase diagram
must be utilized. For example, in ZrZn$_2$ these are the observed
values of $T_{FS}(0)$ and the slope $\rho_0 \equiv [\partial
T_{FS}(P)/\partial P]_0 $ at $P=0$. For UGe$_2$ one may use $T_m$,
$P_m$, and $P_{0c}$, where the last quantity denotes the other
solution (below $P_c$) of $T_{FS}(P)=0$. The ratios
$\gamma/\kappa$ and $\gamma_1/\kappa$ can be deduced using Eq.
(\ref{Eq4}) and the expressions for $T_m$, $P_m$, and $\rho_0$,
while $\kappa$ is chosen by requiring a suitable value of $T_C$.

\begin{figure}
\begin{center}
\epsfig{file=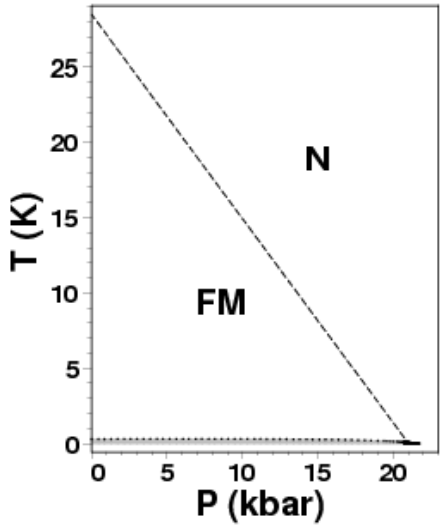,angle=-0, width=6cm}
\end{center}
\caption{$T-P$ diagram of ZrZn$_2$ calculated taking $T_s = 0$,
$\rho_0 = 0.003$ K/kbar, $T_{f0} = 28.5$ K, $P_0 = 21$ kbar,
$\kappa = 10$, $\gamma / \kappa =2\gamma_1 / \kappa \approx 0.2$.
The low-$T$ domain of the FS phase is seen more clearly in the
following figure.} \label{cottam-fig1}
\end{figure}

Experiments \cite{Pfleiderer:2001} for ZrZn$_2$ indicate
$T_{f0} = 28.5$ K, $T_{FS}(0) = 0.29$ K, $P_0 \sim P_c = 21$ kbar,
and $T_F(P)\sim T_f(P)$ is almost a straight line, so $n=1$
describes the $P$-dependence. The slope for $T_{FS}(P)$ at $P=0$
is harder to estimate; its magnitude should not exceed $T_{f0}/P_c
\approx 0.014$ on the basis of a straight-line assumption,
implying $-0.014 <\rho \leq 0$. However, this ignores the effect
of a maximum, although it is unclear experimentally in ZrZn$_2$,
at ($T_m,P_m$). If such a maximum were at $P=0$ we would have
$\rho_0 = 0$, whereas a maximum with $T_m \sim T_{FS}(0)$ and $P_m
\ll P_0$ provides us with an estimated range $0\leq \rho_0 <
0.005$. The choice $\rho_0 = 0$ gives $\gamma / \kappa \approx
0.02$ and $\gamma_1 / \kappa \approx 0.01$, but similar values
hold for any $|\rho_0| \le 0.003$. The multi-critical points A and
C cannot be distinguished experimentally. Since the experimental
accuracy \cite{Pfleiderer:2001} is less than $\sim 25$ mK in the
high-$P$ domain ($P\sim 20-21$ kbar), we suppose that $T_C \sim
10$ mK, which corresponds to $\kappa \sim 10$. We employed these
parameters to calculate the $T-P$ diagram using $\rho_0 = 0$ and
$0.003$. The differences obtained in these two cases are
negligible, with both phase diagrams being in excellent agreement
with experiment.

The latter value is used in Fig. 1, which gives $P_A\sim P_c=
21.10$ kbar, $P_B =20.68$ kbar, $P_C = 20.99$ kbar,
$T_A=T_F(P_c)=T_{FS}(P_c) = 0$ K, $T_B =0.077$ K, $T_C =0.038$ K,
and $T_{FS}(0) =0.285$ K. The low-$T$ region is seen in more
detail in Fig. 2, where the A, B, C points are shown and the order
of the FM-FS phase transition changes from second to first order
around the critical end-point C. The $T_{FS}(P)$ curve has a
maximum at $P_m = 6.915$ kbar and $T_m =0.301$ K. These results
account well for the main features of the experimental behavior
\cite{Pfleiderer:2001}, including the claimed change in the order
of the FM-FS phase transition at relatively high $P$. Within the
present model the N-FM transition is of second order up to $P_C
\sim P_c$. Moreover, if the experiments are reliable in their
indication of a first order N-FM transition at much lower $P$
values, the theory can accommodate this by a change of sign of
$b_f$, leading to a new tricritical point located at a distinct
$P_{tr} < P_C$ on the N-FM transition line. Since $T_C>0$ a direct
N-FS phase transition of first order is predicted in accord with
conclusions from de Haas--van Alphen experiments
\cite{Kimura:2004} and some theoretical studies
\cite{Uhlarz:2004}. Such a transition may not occur in other cases
where $T_C=0$. In SFT ($n=2$) the diagram topology remains the
same but points B and C are slightly shifted to higher $P$
(typically by about 1 bar).

\begin{figure}
\begin{center}
\epsfig{file=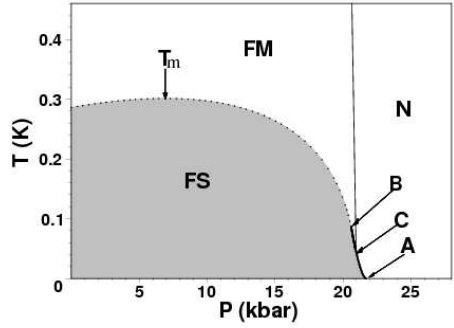,angle=-0, width=6cm}
\end{center}
\caption{Detail of Fig. 1 with expanded temperature scale.}
\label{cottam-fig2}
\end{figure}

The experimental data \cite{Saxena:2000, Tateiwa:2001,
Harada:2007} for UGe$_2$ indicate $T_{f0} = 52$ K, $P_c=1.6$ GPa
($\equiv 16$ kbar), $T_m = 0.75$ K, $P_m\approx 1.15$ GPa, and
$P_{0c} \approx 1.05$ GPa. Using again the variant $n=1$ for
$T_f(P)$ and the above values for $T_m$ and $P_{0c}$ we obtain
$\gamma / \kappa \approx 0.098$ and $\gamma_1 / \kappa \approx
0.168$. The temperature $T_C \sim 0.1$ K corresponds to $\kappa
\sim 5$. Using these, together with $T_s=0$, leads to the $T-P$
diagram in Fig. 3, showing only the low-$T$ region of interest. We
obtain $T_A=0$ K, $T_B=0.481$ K, $T_C=0.301$ K, $P_A=1.72$ GPa,
$P_B = 1.56$ GPa, and $P_C=1.59$ GPa. There is agreement with the
main experimental findings, although $P_m$ corresponding to the
maximum (found at $\sim 1.44$ GPa in Fig. 3) is about 0.3 GPa
higher than suggested experimentally. If the experimental plots
are accurate in this respect, this difference may be attributable
to the so-called ($T_x$) meta-magnetic phase transition in
UGe$_2$, which is related to an abrupt change of the magnetization
in the vicinity of $(T_m,P_m)$. Thus, one may suppose that the
meta-magnetic effects, which are outside the scope of our current
model, significantly affect the shape of the $T_{FS}(P)$ curve by
lowering $P_m$ (along with $P_B$ and $P_C$). It is possible to
achieve a lower $P_m$ value (while leaving $T_m$ unchanged), but
this has the undesirable effect of modifying $P_{c0}$ to a value
that disagrees with experiment. In SFT $(n=2)$ the multi-critical
points are located at slightly higher $P$ (by about 0.01 GPa), as
for ZrZn$_2$.

\begin{figure}
\begin{center}
\epsfig{file=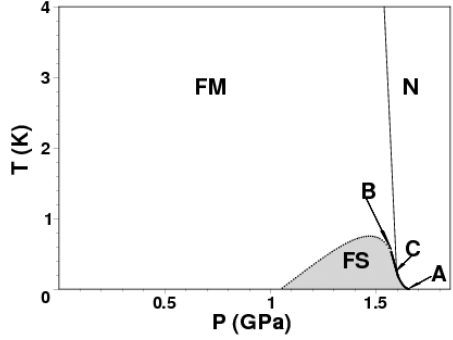,angle=-0, width=6cm}
\end{center}
\caption{Low-$T$ part of the $T-P$ diagram of UGe$_2$ calculated
taking $T_s=0$, $T_{f0}=52$ K, $P_0 = 1.6$ GPa, $T_m = 0.75$ K,
$P_{0c} = 1.05$ GPa ($\kappa = 5$, $\gamma / \kappa = 0.098$, and
$\gamma_1 / \kappa = 0.168$).} \label{cottam-fig3}
\end{figure}

The estimates for UGe$_2$ imply $\gamma_1\kappa\approx 1.9$, so
the condition for $T_{FS}(P)$ to have a maximum found from Eq.
(\ref{Eq4}) is satisfied. As we discussed for ZrZn$_2$, the
location of this maximum can be hard to fix accurately in
experiments. However, $P_{c0}$ can be more easily distinguished,
as in the UGe$_2$ case. Then we have a well-established quantum
(zero-temperature) phase transition of second order, i.e., a
quantum critical point~\cite{Shopova:2003a}. From Eq. (\ref{Eq4})
the existence of this type of solution in systems with $T_s=0$ (as
UGe$_2$) occurs for $\gamma < \gamma_1$. Such systems (which we
label as U-type) are essentially different from those such as
ZrZn$_2$ where $\gamma_1 < \gamma$ and hence $T_{FS}(0) > 0$. In
this latter case (Zr-type compounds) a maximum $T_m> 0$ may
sometimes occur, as discussed earlier. We note that the ratio
$\gamma/\gamma_1$ reflects a balance effect between the two
$\mbox{\boldmath$\psi$}$-$\mbox{\boldmath$M$}$ interactions. When
the trigger interaction (typified by $\gamma$) prevails, the
Zr-type behavior is found where superconductivity exists at $P=0$.
The same ratio can be expressed as $\gamma_0/\delta M_0$, which
emphasizes that the ground state value of the magnetization at
$P=0$ is also relevant. In general, depending on the ratio of the
interaction parameters $\gamma$ and $\gamma_1$, the ferromagnetic
superconductors with spin-triplet Cooper fermion pairing can be of
two types: type I ($\gamma < \gamma_1$) and type II ($\gamma>
\gamma_1$). The two types are distinguished in their thermodynamic
properties.

The quantum phase transition near $P_c$ is of first order.
Depending on the system properties, $T_C$ can be either positive
(when a direct N-FS first order transition is possible), zero, or
negative (when the FM-FS and N-FM phase transition lines terminate
at different zero-temperature phase transition points). The last
two cases correspond to $T_s < 0$. All these cases are possible in
Zr- and U-type compounds. The zero temperature transition at
$P_{c0}$ is found to be a quantum critical point, whereas the
zero-temperature phase transition at $P_c$ is of first order. As
noted, the latter splits into two first order phase transitions.
This classical picture may be changed through quantum fluctuations
\cite{Shopova:2003}. An investigation \cite{Uzunov:2006,
Uzunov:2007} of the quantum fluctuations and the quantum
dimensional crossover by renormalization group methods revealed a
fluctuation change in the order of this transition to a continuous
phase transition belonging to an entirely new class of
universality. However, this option exists only for magnetically
isotropic order (Heisenberg symmetry) and is unlikely to apply in
the known spin-triplet ferromagnetic superconductors.

Even in its simplified form, this theory has been shown to be
capable of accounting for a wide variety of experimental behavior.
A natural extension to the theory is to add a
$\mbox{\boldmath$M$}^6$ term which provides a formalism to
investigate possible metamagnetic phase transitions
\cite{Huxley:2000} and extend some first order phase transition
lines. Another modification of this theory, with regard to
applications to other compounds, is to include a $P$ dependence
for some of the other GL parameters.

Finally, let us emphasize that our results are based on a general
thermodynamic analysis in which the surface and bulk phases are
treated on the same footing. For this reason, our results do not
contradict to experiments~\cite{Yelland:2005} showing a lack of
bulk superconductivity in ZrZn$_2$ but the occurrence of a surface
FS phase at surfaces with higher Zr content than that in ZrZn$_2$.


{\bf Acknowledgements:} The authors are grateful to A. Harada and
S. M. Hayden for valuable communications and discussions of
experimental problems. One of us (D.I.U.) thanks the University of
Western Ontario for hospitality. NFSR-Sofia (through grant Ph.
1507) and NSERC-Canada are acknowledged for partial support.\


\end{document}